\RequirePackage{fix-cm}

\documentclass[smallextended]{svjour3}       % onecolumn (second format)
\usepackage[left=3.17cm,right=3.17cm,top=2.54cm,
headheight=0.5cm,headsep=0.54cm,bottom=2.54cm,footskip=0.79cm
]{geometry}

\usepackage{amsmath,amssymb}

\newtheorem{Thm}{Theorem}

\newtheorem{Cor}{Corollary}
\newtheorem{Exa}{Example}
\usepackage{graphicx}
\begin{document}

\title{Lower bound on concurrence for arbitrary-dimensional tripartite quantum states}%\thanks{Grants or other notes



\author{Wei Chen         \and
        Shao-Ming Fei     \and Zhu-Jun Zheng%etc.
}


\institute{Wei Chen  \at
              School of Automation Science and Engineering, South China University of Technology, Guangzhou 510641, China \\
              \email{auwchen@scut.edu.cn}           %  \\
           \and
           Shao-Ming Fei  \at
              School of Mathematical Sciences, Capital Normal University, Beijing 100048, China
              \and
           Shao-Ming Fei  \at
              Max-Planck-Institute for Mathematics in the Sciences, Leipzig 04103, Germany
              \and
           Zhu-Jun Zheng  \at
           School of Mathematics, South China University of Technology,
Guangzhou 510641, China
}

\date{Received: date / Accepted: date}

\maketitle

\begin{abstract}
In this paper, we study the concurrence of arbitrary-dimensional tripartite quantum states. An explicit operational lower bound of concurrence is obtained in terms of the concurrence of substates.  A given example show that our lower bound may improve the well known existing lower bounds of concurrence. The significance of our result is to get a lower bound when we study the concurrence of arbitrary $m\otimes n\otimes l$-dimensional tripartite quantum states.
 %\PACS{03.67.Mn \and 03.65.Ud }
\end{abstract}

\section{Introduction}
\label{intro}
As one of the most striking features of quantum phenomena\cite{Per}, quantum entanglement has been identified as a key non-local resource in quantum information processing vary from quantum teleportation\cite{Ben} and quantum cryptography\cite{Eke} to dense coding\cite{Ben1}. These effects based on quantum entanglement have been demonstrated in many outstanding experiments.

An important issue in the theory of quantum entanglement is to recognize and quantify the entanglement for a given quantum state. Concurrence is one of the well-defined quantitative measures of entanglement\cite{Uhl}-\cite{Woo}. For a mixed two-qubit state, an elegant formula of concurrence was derived analytically by Wootters in \cite{Woo}. However, beyond bipartite qubit systems and some special symmetric states \cite{Ter}, there exists no explicit analytic formulas to show the concurrence of arbitrary high-dimensional mixed states. Instead of analytic formulas, some progress has been made toward the analytical lower bounds of concurrence. In recent years, there are many papers \cite{Flo}-\cite{Min} to give the lower bounds of concurrence for bipartite quantum states by using different methods. All these bounds give rise to a good quantitative estimation of concurrence. They are usually supplementary in detecting quantum entanglement.

With the deepening research of the lower bound of bipartite concurrence, some nice algorithms and progress have been concentrated on possible lower bound of concurrence for tripartite quantum systems \cite{Gao}-\cite{Zhu3} and other multipartite quantum systems \cite{Wang}-\cite{Zhu2}. In Ref.\cite{Wang,Zhu1}, the authors obtained lower bounds of multipartite concurrence in terms of the concurrence of bipartite partitioned states of the whole quantum system. In Ref.\cite{Chen}, a lower bound of concurrence for four-qubit systems has been presented by using tripartite concurrence. However, the lower bound (10) in \cite{Chen} is in general not operationally computable. In order to get operationally lower bound for multipartite quantum systems, we need to give operationally lower bounds for $m\otimes n \otimes l$ $(m\neq n\neq l)$ tripartite quantum systems, which will be considered in this paper.

In Ref.\cite{Gao}, the analytic lower bounds of concurrence for three-qubit systems or for any $m\otimes n \otimes l$ $(m\leq n, l)$ tripartite quantum systems have been presented by using the generalized partial transposition(GPT) criterion. In Ref.\cite{Zhu3}, another analytic lower bounds of concurrence for $N\otimes N \otimes N$ tripartite quantum systems have been obtained in terms of the concurrence of substates.

This paper is organized as follows. In Sec. 2, we generalize the results in \cite{Gao,Zhu3} and obtain some new operational lower bounds of concurrence for arbitrary $m\otimes n \otimes l$-dimensional $(m\neq n\neq l)$ tripartite quantum systems in terms of lower-dimensional systems. In Sec. 3, we present that our lower bound may be used to improve the known lower bounds of concurrence with an example. Conclusions are given in Sec. 4.

\section{Lower bounds of concurrence for tripartite quantum systems of different dimensions}
\label{sec:1}
We first recall the definition of the tripartite concurrence. Let $H_{A_1}, H_{A_2}$ and $H_{A_3}$ be $m$-, $n$-, $l$-dimensional Hilbert spaces, respectively. In general, we can assume that $m\leq n \leq l,$ any pure tripartite state $|\psi\rangle\in H_{A_1}\otimes H_{A_2}\otimes H_{A_3}$ has the form
\begin{equation}
|\psi\rangle = \sum_{i=1}^{m}\sum_{j=1}^{n}\sum_{k=1}^{l}a_{ijk}|ijk\rangle,
\end{equation}
where $a_{ijk}\in \mathbb{C}, \sum_{ijk}|a_{ijk}|^{2}=1, \{|ijk\rangle\}$ is the basis of $ H_{A_1}\otimes H_{A_2}\otimes H_{A_3}.$

The concurrence of a tripartite pure state $|\psi\rangle\in H_{A_1}\otimes H_{A_2}\otimes H_{A_3}$ is defined by \cite{Aol}

\begin{equation}
C(|\psi\rangle) = \sqrt{3- Tr(\rho_{A_1}^{2}+\rho_{A_2}^{2}+\rho_{A_3}^{2})},
\end{equation}
where the reduced density matrix $\rho_{A_1}$ (respectively, $\rho_{A_2}, \rho_{A_3}$) is obtained by tracing over the subsystems $A_2$ and $A_3$(respectively, $A_1$ and $A_3$, $A_1$ and $A_2$). When $m=n=l,$ $C(|\psi\rangle)$ can be equivalently written as \cite{Alb}

\begin{equation}
C(|\psi\rangle) = \sqrt{\frac{1}{2}\sum(|a_{ijk}a_{pqt}-a_{ijt}a_{pqk}|^2+|a_{ijk}a_{pqt}-a_{iqk}a_{pjt}|^2+|a_{ijk}a_{pqt}-a_{pjk}a_{iqt}|^2)}.
\end{equation}

When $m\neq n\neq l,$ we can have the following result:

\begin{Thm}
For any $m, n, l,$ we have
\begin{equation}
C^{2}(|\psi\rangle) = \frac{1}{2}\sum^{m}_{i,p=1}\sum^{n}_{j,q=1}\sum^{l}_{k,t=1}(|a_{ijk}a_{pqt}-a_{ijt}a_{pqk}|^2+|a_{ijk}a_{pqt}-a_{iqk}a_{pjt}|^2+|a_{ijk}a_{pqt}-a_{pjk}a_{iqt}|^2).
\end{equation}
\end{Thm}
Proof. For any $m, n, l,$ a pure tripartite state $|\psi\rangle\in H_{A_1}\otimes H_{A_2}\otimes H_{A_3}$ has the form
$$|\psi\rangle = \sum_{i=1}^{m}\sum_{j=1}^{n}\sum_{k=1}^{l}a_{ijk}|ijk\rangle,$$ then we can compute
$$\rho_{A_1}=\sum^{m}_{i,p=1}\sum^{n}_{j=1}\sum^{l}_{k=1}a_{ijk}a^{*}_{pjk}|i\rangle\langle p|,$$ and
$$tr\rho^{2}_{A_1}=\sum^{m}_{i,p=1}\sum^{n}_{j,q=1}\sum^{l}_{k,t=1}a_{ijk}a^{*}_{pjk}a_{pqt}a^{*}_{iqt},$$ hence we obtain
$$1-tr\rho^{2}_{A_1}=\frac{1}{2}\sum^{m}_{i,p=1}\sum^{n}_{j,q=1}\sum^{l}_{k,t=1}|a_{ijk}a_{pqt}-a_{pjk}a_{iqt}|^2.$$
Similarly, we have $$1-tr\rho^{2}_{A_2}=\frac{1}{2}\sum^{m}_{i,p=1}\sum^{n}_{j,q=1}\sum^{l}_{k,t=1}|a_{ijk}a_{pqt}-a_{iqk}a_{pjt}|^2,$$ and
$$1-tr\rho^{2}_{A_3}=\frac{1}{2}\sum^{m}_{i,p=1}\sum^{n}_{j,q=1}\sum^{l}_{k,t=1}|a_{ijk}a_{pqt}-a_{ijt}a_{pqk}|^2.$$
Associated with (2), we get our result (4).

The concurrence for a tripartite mixed state $\rho$ is defined by the convex roof,

\begin{equation}
C_N(\rho) = \min_{\{p_{i},|\psi_{i}\rangle\}}\sum_{i}p_{i}C(|\psi_{i}\rangle),
\end{equation}
where the minimum is taken over all possible convex decompositions of $\rho$ into an ensemble $\{|\psi_{i}\rangle\}$ of pure states with probability distribution $\{p_{i}\}.$

To evaluate $C(\rho),$ we project high-dimensional states to "lower-dimensional" ones. For a given $m\otimes n \otimes l$ pure state $|\psi\rangle$, we define its $s_1\otimes s_2 \otimes s_3, s_1\leq m, s_2\leq n, s_3\leq l$ pure substate $|\psi\rangle_{s_1\otimes s_2 \otimes s_3}=\sum^{i_{s_1}}_{i=i_{1}}\sum^{j_{s_2}}_{j=j_{1}}\sum^{k_{s_3}}_{k=k_{1}}a_{ijk}|ijk\rangle=E_1\otimes E_2 \otimes E_3|\psi\rangle,$
where $E_1=\sum^{i_{s_1}}_{i=i_{1}}|i\rangle\langle i|, E_2=\sum^{j_{s_2}}_{j=j_{1}}|j\rangle\langle j|$ and $E_3=\sum^{k_{s_3}}_{k=k_{1}}|k\rangle\langle k|.$
We denote the concurrence $C^{2}(|\psi\rangle_{s_1\otimes s_2 \otimes s_3})$ by
$$C^{2}(|\psi\rangle_{s_1\otimes s_2 \otimes s_3})= \frac{1}{2}\sum_{u,x=1}^{s_1}\sum_{v,y=1}^{s_2}\sum_{w,z=1}^{s_3}|a_{i_uj_vk_w}a_{p_xq_yt_z}-a_{p_xj_vk_w}a_{i_uq_yt_z}|^2$$
$$~~~~~~~~~~~~~~~~~~~~+\frac{1}{2}\sum_{u,x=1}^{s_1}\sum_{v,y=1}^{s_2}\sum_{w,z=1}^{s_3}|a_{i_uj_vk_w}a_{p_xq_yt_z}-a_{i_uq_yk_w}a_{p_xj_vt_z}|^2$$
$$~~~~~~~~~~~~~~~~~~~~+\frac{1}{2}\sum_{u,x=1}^{s_1}\sum_{v,y=1}^{s_2}\sum_{w,z=1}^{s_3}|a_{i_uj_vk_w}a_{p_xq_yt_z}-a_{i_uj_vt_z}a_{p_xq_yk_w}|^2.$$
In fact, there are $\binom{m}{s_1}\times \binom{n}{s_2} \times\binom{l}{s_3}$ different $s_1\otimes s_2 \otimes s_3$ substates $|\psi\rangle_{s_1\otimes s_2 \otimes s_3}$ for a given pure state $|\psi\rangle,$ where  $\binom{m}{s_1}, \binom{n}{s_2}$ and $\binom{l}{s_3}$ are the binomial coefficients. To avoid causing confusion, in the following we simply use $|\psi\rangle_{s_1\otimes s_2 \otimes s_3}$ to denote one of such substates, as these substates will always be considered together.

For a mixed state $\rho \in H_{A_1}\otimes H_{A_2}\otimes H_{A_3},$ we define its $s_1\otimes s_2 \otimes s_3$ mixed substates by $\rho_{s_1\otimes s_2 \otimes s_3} = E_1\otimes E_2 \otimes E_3 \rho E_1^{\dagger}\otimes E_2^{\dagger} \otimes E_3^{\dagger},$  having the following matrices form:

{\begin{equation}
\setlength{\arraycolsep}{0.2pt}
\rho_{s_1\otimes s_2 \otimes s_3}=
\left (
\begin{array}{cccccccc}
\rho_{i_1j_1k_1,i_1j_1k_1}&\cdots&\rho_{i_1j_1k_1,i_1j_1k_{s_3}}&\rho_{i_1j_1k_1,i_1j_2k_1}&\cdots&\rho_{i_1j_1k_1,i_{s_1}j_{s_2}k_{s_3}}\\
\vdots&\vdots&\vdots&\vdots&\vdots&\vdots&\\
\rho_{i_1j_1k_{s_3},i_1j_1k_1}&\cdots&\rho_{i_1j_1k_{s_3},i_1j_1k_{s_3}}&\rho_{i_1j_1k_{s_3},i_1j_2k_1}&\cdots&\rho_{i_1j_1k_{s_3},i_{s_1}j_{s_2}k_{s_3}}\\
\rho_{i_1j_2k_1,i_1j_1k_1}&\cdots&\rho_{i_1j_2k_1,i_1j_1k_{s_3}}&\rho_{i_1j_2k_1,i_1j_2k_1}&\cdots&\rho_{i_1j_2k_1,i_{s_1}j_{s_2}k_{s_3}}\\
\vdots&\vdots&\vdots&\vdots&\vdots&\vdots&\\
\rho_{i_1j_2k_{s_3},i_1j_1k_1}&\cdots&\rho_{i_1j_2k_{s_3},i_1j_1k_{s_3}}&\rho_{i_1j_2k_{s_3},i_1j_2k_1}&\cdots&\rho_{i_1j_2k_{s_3},i_{s_1}j_{s_2}k_{s_3}}\\
%\vdots&\vdots&\vdots&\vdots&\vdots&\vdots&\vdots&\vdots\\
%\rho_{i_1j_{s_2}k_1,i_1j_1k_1}&\cdots&\rho_{i_1j_{s_2}k_1,i_1j_1k_{s_3}}&\rho_{i_1j_{s_2}k_1,i_1j_2k_1}&\cdots&\rho_{i_1j_{s_2}k_1,i_1j_2k_{s_3}}&\cdots&\rho_{i_1j_{s_2}k_1,i_{s_1}j_{s_2}k_{s_3}}\\
%\vdots&\vdots&\vdots&\vdots&\vdots&\vdots&\vdots&\vdots\\
%\rho_{i_1j_{s_2}k_{s_3},i_1j_1k_1}&\cdots&\rho_{i_1j_{s_2}k_{s_3},i_1j_1k_{s_3}}&\rho_{i_1j_{s_2}k_{s_3},i_1j_2k_1}&\cdots&\rho_{i_1j_{s_2}k_{s_3},i_1j_2k_{s_3}}&\cdots&\rho_{i_1j_{s_2}k_{s_3},i_{s_1}j_{s_2}k_{s_3}}\\
\vdots&\vdots&\vdots&\vdots&\vdots&\vdots&\\
\rho_{i_{s_1}j_{s_2}k_{s_3},i_1j_1k_1}&\cdots&\rho_{i_{s_1}j_{s_2}k_{s_3},i_1j_1k_{s_3}}&\rho_{i_{s_1}j_{s_2}k_{s_3},i_1j_2k_1}&\cdots&\rho_{i_{s_1}j_{s_2}k_{s_3},i_{s_1}j_{s_2}k_{s_3}}\\
\end{array}
\right ),
\end{equation}
}which are unnormalized tripartite $s_1\otimes s_2 \otimes s_3$ mixed states. The concurrence of $\rho_{s_1\otimes s_2 \otimes s_3}$ is defined by $C(\rho_{s_1\otimes s_2 \otimes s_3})\equiv min\sum_{i}p_iC(|\psi_i\rangle_{s_1\otimes s_2 \otimes s_3}),$ minimized over all possible $s_1\otimes s_2 \otimes s_3$ pure-state decompositions of $\rho_{s_1\otimes s_2 \otimes s_3}=\sum_{i}p_i|\psi_i\rangle_{s_1\otimes s_2 \otimes s_3}\langle\psi_i|,$ with $\sum_{i}p_i=Tr(\rho_{s_1\otimes s_2 \otimes s_3}).$

\begin{Thm}
For any $m\otimes n\otimes l$ tripartite mixed quantum state $\rho\in H_{A_1}\otimes H_{A_2}\otimes H_{A_3},$ assume $2\leq m\leq n\leq l,$ then the concurrence $C(\rho)$ satisfies

\begin{equation}
C^2(\rho)\geq c_{s\otimes s\otimes s}\sum_{P_{s\otimes s\otimes s}}C^2(\rho_{s\otimes s\otimes s})\equiv\tau_{s\otimes s\otimes s}(\rho),
\end{equation}
where $m\geq s\geq 2, c_{s\otimes s\otimes s}= [\binom{m-2}{s-2}\times \binom{n-2}{s-2}\times \binom{l-1}{s-1}]^{-1}, \sum_{P_{s\otimes s\otimes s}}$ stands for
summing over all possible $s\otimes s\otimes s$ mixed substates, and $\tau_{s\otimes s\otimes s}(\rho)$ denotes the lower bound of $C^2(\rho)$ with respect to the $s\otimes s\otimes s$ subspaces.
\end{Thm}
Proof. For any $m\otimes n\otimes l$ tripartite pure quantum state$|\psi\rangle = \sum_{i=1}^{m}\sum_{j=1}^{n}\sum_{k=1}^{l}a_{ijk}|ijk\rangle,$ and any given term
\begin{equation}
|a_{i_0j_0k_0}a_{p_0q_0t_0}-a_{i_0j_0t_0}a_{p_0q_0k_0}|^2, i_0\neq p_0,
\end{equation}
in Eq.(4).

If $j_0\neq q_0$ and $k_0\neq t_0,$ then there are $\binom{m-2}{s-2}\times \binom{n-2}{s-2}\times \binom{l-2}{s-2}$ different $s\otimes s\otimes s$ substates  $|\psi\rangle_{s\otimes s \otimes s}=E_1\otimes E_2 \otimes E_3|\psi\rangle,$ with $E_1=|i_0\rangle \langle i_0| + |p_0\rangle \langle p_0| + \sum^{i_{s}}_{i=i_{3}}|i\rangle\langle i|, E_2=|j_0\rangle \langle j_0|+|q_0\rangle \langle q_0|+\sum^{j_{s}}_{j=j_{3}}|j\rangle\langle j|, E_3=|k_0\rangle \langle k_0|+|t_0\rangle \langle t_0|+\sum^{k_{s}}_{k=k_{1}}|k\rangle\langle k|,$ where $\{|i\rangle \}_{i=i_3}^{i_s}\subseteq \{|i\rangle \}_{i=1}^{m},$  $\{|j\rangle \}_{j=j_3}^{j_s}\subseteq \{|j\rangle \}_{j=1}^{n}$ and  $\{|k\rangle \}_{k=k_3}^{k_s}\subseteq \{|k\rangle \}_{k=1}^{l},$ such that the term (8) appears in the concurrence of $|\psi\rangle_{s\otimes s \otimes s}=E_1\otimes E_2 \otimes E_3|\psi\rangle.$

If $j_0\neq q_0$ and $k_0 = t_0,$ then there are $\binom{m-2}{s-2}\times \binom{n-2}{s-2}\times \binom{l-1}{s-1}$ different $s\otimes s\otimes s$ substates  $|\psi\rangle_{s\otimes s \otimes s}=E_1\otimes E_2 \otimes F_3|\psi\rangle,$ with $F_3=|k_0\rangle \langle t_0|  + \sum^{k_{s}}_{k=k_{2}}|k\rangle\langle k|,$ where $\{|k\rangle \}_{k=k_2}^{k_s}\subseteq \{|k\rangle \}_{k=1}^{m},$  such that the term (8) appears in the concurrence of $|\psi\rangle_{s\otimes s \otimes s}=E_1\otimes E_2 \otimes F_3|\psi\rangle.$

Otherwise, if $j_0 = q_0$ and $k_0\neq t_0,$ then there are $\binom{m-2}{s-2}\times \binom{n-1}{s-1}\times \binom{l-2}{s-2}$ different $s\otimes s\otimes s$ substates  $|\psi\rangle_{s\otimes s \otimes s}=E_1\otimes F_2 \otimes E_3|\psi\rangle,$ with $F_2=|j_0\rangle \langle j_0|  + \sum^{j_{s}}_{j=j_{2}}|j\rangle\langle j|,$ where $\{|j\rangle \}_{j=j_2}^{j_s}\subseteq \{|j\rangle \}_{j=1}^{n},$  such that the term (8) appears in the concurrence of $|\psi\rangle_{s\otimes s \otimes s}=E_1\otimes F_2 \otimes E_3|\psi\rangle.$

Since $\binom{l-2}{s-2}\leq \binom{l-1}{s-1}$ and $\binom{n-1}{s-1}\times \binom{l-2}{s-2} \leq \binom{n-2}{s-2}\times \binom{l-1}{s-1},$ we have the following relation:
\begin{equation}
\binom{m-2}{s-2}\times \binom{n-2}{s-2}\times \binom{l-1}{s-1}C^2(|\psi\rangle)\geq \sum_{P_{s\otimes s\otimes s}}C^2(|\psi\rangle_{s\otimes s\otimes s})
\end{equation}
equivalently,

\begin{equation}
C^2(|\psi\rangle)\geq c_{s\otimes s\otimes s}\sum_{P_{s\otimes s\otimes s}}C^2(|\psi\rangle_{s\otimes s\otimes s})
\end{equation}

Therefore, for mixed state $\rho=\sum p_i|\psi_i\rangle\langle \psi_i|,$ we have
$$C(\rho)=min\sum_i p_iC(|\psi_i\rangle)~~~~~~~~~~~~~~$$
$$~~~~~~~~~~~~~~~~~~~~~\geq \sqrt{c_{s\otimes s\otimes s}}min\sum_i p_i(\sum_{P_{s\otimes s\otimes s}}C^2(|\psi_i\rangle_{s\otimes s\otimes s}))^{\frac{1}{2}}$$
$$~~~~~~~~~~~~~~~~~~~~~\geq \sqrt{c_{s\otimes s\otimes s}}min[\sum_{P_{s\otimes s\otimes s}}(\sum_i p_iC(|\psi_i\rangle_{s\otimes s\otimes s}))^2]^{\frac{1}{2}}$$
$$~~~~~~~~~~~~~~~~~~~~~\geq \sqrt{c_{s\otimes s\otimes s}}[\sum_{P_{s\otimes s\otimes s}}(min\sum_i p_iC(|\psi_i\rangle_{s\otimes s\otimes s}))^2]^{\frac{1}{2}}$$
$$~~~~~~~~~~~~~~~~~~~~~=\sqrt{c_{s\otimes s\otimes s}}[\sum_{P_{s\otimes s\otimes s}}C^2(\rho_{s\otimes s\otimes s})]^{\frac{1}{2}},~~~~~~~~~~~~~~~~$$
where the relation $[\sum_j(\sum_ix_{ij})^2]^{\frac{1}{2}}\leq \sum_i(\sum_jx_{ij}^2)^{\frac{1}{2}}$ has been used in the second inequality, the first three minimizations run over all possible pure-state decompositions of the mixed state $\rho,$ while the last minimization runs over all $s\otimes s\otimes s$ pure-state decompositions of $\rho_{s\otimes s\otimes s}=\sum_i p_i|\psi_i\rangle\langle \psi_i|$ associated with $\rho.$

Equation (7) gives a lower bound of $C(\rho).$ One can estimate $C(\rho)$ by calculating the concurrence of the substates $\rho_{s\otimes s\otimes s}, 2\leq s < m.$ Different choices of $s$ may give rise to different lower bounds. A convex combination of these lower bounds is still a lower bound. So, generally we have the following:

\begin{Cor}

For any $m\otimes n\otimes l$ tripartite mixed quantum state $\rho\in H_{A_1}\otimes H_{A_2}\otimes H_{A_3}, s\geq 2,$ assume $2\leq m\leq n\leq l,$ then the concurrence $C(\rho)$ satisfies

\begin{equation}
C^2(\rho)\geq \sum_{s=2}^{m}p_s \tau_{s\otimes s\otimes s}(\rho),
\end{equation}
where $0\leq p_s \leq 1, s=2, \cdot\cdot\cdot, m$ and $\sum_{s=2}^{m}p_s=1.$

\end{Cor}

\begin{Thm}
For any $s\otimes s\otimes s$ tripartite mixed quantum state $\rho\in H_{A_1}\otimes H_{A_2}\otimes H_{A_3}, s\geq 2,$ then the concurrence $C(\rho)$ satisfies

\begin{equation}
C^2(\rho)\geq c_{\lambda\otimes \mu \otimes \nu}\sum_{P_{\lambda\otimes \mu\otimes \nu}}C^2(\rho_{\lambda\otimes \mu\otimes \nu})\equiv\tau_{\lambda\otimes \mu\otimes \nu}(\rho),
\end{equation}
where $1<\lambda\leq \mu \leq \nu\leq s, c_{\lambda\otimes \mu\otimes \nu}= [\binom{s-1}{\lambda-1}\times \binom{s-2}{\mu-2}\times \binom{s-2}{\nu-2}]^{-1}, \sum_{P_{\lambda\otimes \mu\otimes \nu}}$ stands for
summing over all possible $\lambda\otimes \mu\otimes \nu$ mixed substates, and $\tau_{\lambda\otimes \mu\otimes \nu}(\rho)$ denotes the lower bound of $C^2(\rho)$ with respect to the $\lambda\otimes \mu\otimes \nu$ subspace.
\end{Thm}
Proof. The proof is as same as Theorem 2.

The lower bound of concurrence of $\rho$ in equation (12) is given by the concurrence of sub-matrix $\rho_{\lambda\otimes \mu\otimes \nu}.$ Choosing different $\lambda, \mu$ and $\nu$ would result in different lower bounds. Generally, we have the following corollary.

\begin{Cor}

For any $s\otimes s\otimes s$ tripartite mixed quantum state $\rho\in H_{A_1}\otimes H_{A_2}\otimes H_{A_3}, s\geq 2,$ then the concurrence $C(\rho)$ satisfies

\begin{equation}
C^2(\rho)\geq \sum^s_{\lambda=2}\sum^s_{\mu\geq\lambda}\sum^s_{\nu\geq\mu}p_{\lambda\mu\nu}\tau_{\lambda\otimes \mu\otimes \nu}(\rho),
\end{equation}
where $0\leq p_{\lambda\mu\nu} \leq 1, 1<\lambda\leq \mu \leq \nu\leq s$ and $\sum^s_{\lambda=2}\sum^s_{\mu\geq\lambda}\sum^s_{\nu\geq\mu}p_{\lambda\mu\nu}=1.$

\end{Cor}

\section{Lower bounds of concurrence for tripartite quantum systems from lower bounds}
\label{sec:2}
The lower bounds (7) and (12) are in general not operationally computable, as we still have no analytical results for concurrence of lower-dimensional states.  If we replace the computation of concurrence of lower-dimensional substates $\rho_{s\otimes s\otimes s}$ and $\rho_{\lambda\otimes \mu\otimes \nu}$ by that of the lower bounds of  three-qubit mixed quantum substates, Eq.(7) and (12) give an operational lower bound based on known lower bounds. The lower bound obtained in this way should be the same or better than the previously known lower bounds. Hence (7) and (12) can be used to improve all the known lower bounds of concurrence by associating with some analytical lower bounds for three-qubit mixed quantum states  \cite{Gao,Zhu3} in this sense.

Assume $g(\rho)$ is any lower bound of concurrence, i.e. $C(\rho)\geq g(\rho).$ Then for a given mixed state $\rho,$ the concurrence of the projected lower-dimensional mixed state $\rho_{s\otimes s\otimes s}$ satisfies
\begin{equation}
C(\rho_{s\otimes s\otimes s}) = tr(\rho_{s\otimes s\otimes s})C((tr\rho_{s\otimes s\otimes s})^{-1}\rho_{s\otimes s\otimes s})\geq  tr(\rho_{s\otimes s\otimes s})g((tr\rho_{s\otimes s\otimes s})^{-1}\rho_{s\otimes s\otimes s}).
\end{equation}
Associated with (7), we get
\begin{equation}
C^2(\rho)\geq c_{s\otimes s\otimes s}\sum_{P_{s\otimes s\otimes s}}C^2(\rho_{s\otimes s\otimes s})\geq c_{s\otimes s\otimes s}\sum_{P_{s\otimes s\otimes s}}(tr(\rho_{s\otimes s\otimes s}))^{2}g^{2}((tr\rho_{s\otimes s\otimes s})^{-1}\rho_{s\otimes s\otimes s}).
\end{equation}
Here if we choose $\rho_{s\otimes s\otimes s}$ to be the given mixed state $\rho$ itself, the inequality reduces to $C(\rho)\geq g(\rho)$ again. Generally, the lower bound $g(\rho)$ may be improved if one takes into account all the lower-dimensional mixed states $\rho_{s\otimes s\otimes s}.$

In the following, we will first present an analytical lower bound for $2\otimes 2\otimes 2$ mixed quantum substates like $\rho_{2\otimes 2\otimes 2}$ by using the Theorem 2 in \cite{Zhu3} and (15).

\begin{Thm}
For any $m\otimes n\otimes l$ tripartite mixed quantum state $\rho\in H_{A_1}\otimes H_{A_2}\otimes H_{A_3},$ assume $2\leq m\leq n\leq l,$ let $\rho_{2\otimes 2\otimes 2}$ be a $2\otimes 2\otimes 2$  mixed quantum sub-state, then we have

\begin{equation}
C^2(\rho_{2\otimes 2\otimes 2})\geq \frac{1}{2}max[\sum^{3}_{j=1}(||\rho_{2\otimes 2\otimes 2}^{\mathcal{T}_j}||-tr(\rho_{2\otimes 2\otimes 2}))^{2}, \sum^{3}_{j=1}(||R_{j,\bar{j}}(\rho_{2\otimes 2\otimes 2})||-tr(\rho_{2\otimes 2\otimes 2}))^{2}],
\end{equation}
and
\begin{equation}
C^2(\rho)\geq \frac{1}{l-1}\sum_{P_{2\otimes 2\otimes 2}}\frac{1}{2}max[\sum^{3}_{j=1}(||\rho_{2\otimes 2\otimes 2}^{\mathcal{T}_j}||-tr(\rho_{2\otimes 2\otimes 2}))^{2}, \sum^{3}_{j=1}(||R_{j,\bar{j}}(\rho_{2\otimes 2\otimes 2})||-tr(\rho_{2\otimes 2\otimes 2}))^{2}]
\end{equation}
where $\rho_{2\otimes 2\otimes 2}^{\mathcal{T}_j}$ stands for partial transposition of $\rho_{2\otimes 2\otimes 2}$ with respect to the $j$th sub-system $A_j,$ $R_{j,\bar{j}}(\rho_{2\otimes 2\otimes 2})$ is the realignment of $\rho_{2\otimes 2\otimes 2}$ with respect to the bipartite partition between $j$th and the rest systems, and $||A||=Tr\sqrt{AA^{\dag}}$ is the trace norm of a matrix.
\end{Thm}
Proof. For the three-qubit mixed quantum state $\rho$, by Theorem 2 in \cite{Zhu3}, we have $$C^2(\rho)\geq \frac{1}{2}max[\sum^{3}_{j=1}(||\rho^{\mathcal{T}_j}||-1)^{2}, \sum^{3}_{j=1}(||R_{j,\bar{j}}(\rho)||-1)^{2}]=:g^2(\rho).$$

From (14), we get
$$C^{2}(\rho_{s\otimes s\otimes s}) \geq  (tr(\rho_{s\otimes s\otimes s}))^2g^2((tr\rho_{s\otimes s\otimes s})^{-1}\rho_{s\otimes s\otimes s})~~~~~~~~~~~~~~~~~~~~~~~~~~~~~~~~~~~~~~~~~~~~~~~~~~~$$
$$~\geq  (tr(\rho_{s\otimes s\otimes s}))^2\frac{1}{2}max[\sum^{3}_{j=1}(||(tr\rho_{s\otimes s\otimes s})^{-1}\rho_{s\otimes s\otimes s}^{\mathcal{T}_j}||-1)^{2}, \sum^{3}_{j=1}(||R_{j,\bar{j}}((tr\rho_{s\otimes s\otimes s})^{-1}\rho_{s\otimes s\otimes s})||-1)^{2}]$$
$$=\frac{1}{2}max[\sum^{3}_{j=1}(||\rho_{2\otimes 2\otimes 2}^{\mathcal{T}_j}||-tr(\rho_{2\otimes 2\otimes 2}))^{2}, \sum^{3}_{j=1}(||R_{j,\bar{j}}(\rho_{2\otimes 2\otimes 2})||-tr(\rho_{2\otimes 2\otimes 2}))^{2}].~~~~~~~~~~~~~~~~~~~~~~~~~~$$
Then, associated with (15), we obtain (17).

(17) gives an operational lower bound of concurrence for any $m\otimes n\otimes l$ tripartite mixed quantum state. We will show the power of (17)
by the following example:

\begin{Exa}
We consider the $2\otimes 2\otimes 4$ quantum mixed state $\rho = \frac{1-t}{16}I_{16} + t|\phi\rangle \langle \phi|,$ with $|\phi\rangle = \frac{1}{2}(|000\rangle + |003\rangle + |110\rangle + |113\rangle),$
where $0\leq t \leq 1$ and $I_{16}$ denotes the $16\times 16$ identity matrix. According to (17), we obtain
\[C^2(\rho)\geq \left\{\begin{array}{ll}
~~~~~~0,&\text{$0\leq t \leq\frac{1}{9}$ },\\
\frac{81t^{2}-18t+1}{96},&\text{$\frac{1}{9}< t \leq\frac{1}{5}$},\\
\frac{181t^{2}-58t+5}{96},&
\text{$\frac{1}{5}< t \leq 1$}.
\end{array}\right.\]
So our result can detect the entanglement of $\rho$ when
$\frac{1}{9}< t \leq 1,$ see Fig.1. While the lower bound of Theorem 2 in \cite{Gao} is $C^2(\rho)\geq 0,$ which can not detect the entanglement of the above $\rho.$

\begin{figure}[htpb]\label{fig4}
\centering\
\includegraphics[width=10cm]{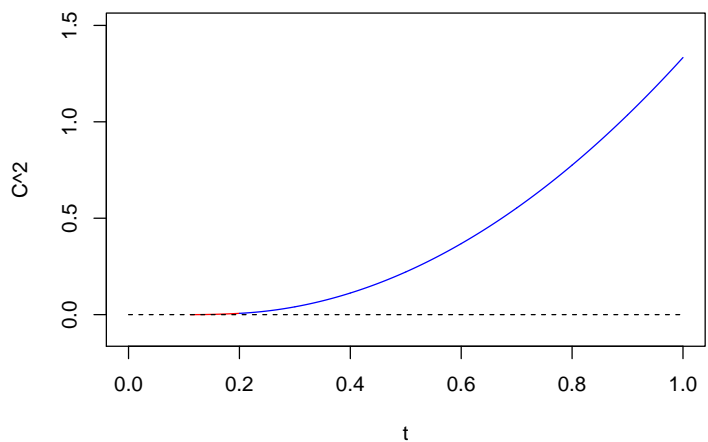}
\caption{Color line for the lower bound of $\rho$ for $\frac{1}{9}< t \leq 1$ from (17), dashed line for the lower bound from Theorem 2 in \cite{Gao}.}
\end{figure}

\end{Exa}

\section{Conclusion}\label{sec3}

In summary, we have perform a method of constructing some new lower bounds of concurrence for tripartite mixed states in terms of the concurrence of substates. By an example we have shown that this bound is better for some states than other existing lower bounds of concurrence. Also the approach can be readily generalized to arbitrary dimensional multipartite systems.

\section*{Acknowledgements}
This project is supported by NSFC through Grants No. 11571119, 11405060, 11475178 and 11275131.

%\begin{acknowledgements}
%\end{acknowledgements}



\end{document}